\documentstyle[12pt]{article}
\def\a{\alpha}

\def\d{\delta}

\def\h{\eta}

\def\l{\lambda}
\def\m{\mu}

\def\p{\pi}       
\def\q{\theta}    

\def\D{\Delta}

\def\O{\Omega}

\def\mh{\hat{\m}}

\newcommand{\ncm}{\newcommand}
\ncm{\rencm}{\renewcommand}
\ncm{\dsp}{\displaystyle}
\ncm{\nn}{\nonumber}
\ncm{\nit}{\noindent}
\ncm{\av}[1]{\mbox{$\langle #1 \rangle$}}
\ncm{\avc}[1]{\mbox{$\langle #1 \rangle_{\psi}$}}
\ncm{\Pm}{\mbox{$m_{P}$}}
\ncm{\Pmt}{\mbox{$m_{P}^2$}}
\ncm{\Pmti}{\mbox{$m_{P}^{-2}$}}
\ncm{\Pmf}{\mbox{$m_{P}^4$}}
\ncm{\half}{\mbox{{\small $\frac{1}{2}$}} }
\ncm{\quart}{\mbox{{\small $\frac{1}{4}$}} }
\ncm{\tq}{\mbox{{\small $\frac{3}{4}$}} }
\ncm{\third}{\mbox{{\small $\frac{1}{3}$}} }
\ncm{\sixth}{\mbox{{\small $\frac{1}{6}$}} }
\ncm{\eigth}{\mbox{{\small $\frac{1}{8}$}} }
\ncm{\thrhalf}{\mbox{{\small $\frac{3}{2}$}} }
\ncm{\thrfor}{\mbox{{\small $\frac{3}{4}$}} }
\ncm{\twothi}{\mbox{{\small $\frac{2}{3}$}} }
\ncm{\fivtwo}{\mbox{{\small $\frac{5}{2}$}} }
\ncm{\ninhalf}{\mbox{{\small $\frac{9}{2}$}} }
\ncm{\ninth}{\mbox{{\small $\frac{1}{9}$}} }
\ncm{\nist}{\mbox{{\small $\frac{9}{16}$}} }
\ncm{\df}{\mbox{$\partial_{\phi}$}}
\ncm{\dft}{\mbox{$\partial_{\phi}^2$}}
\ncm{\da}{\mbox{$\partial_{a}$}}
\ncm{\dat}{\mbox{$\partial_{a}^2$}}
\ncm{\dath}{\mbox{$\partial_{a}^3$}}
\ncm{\dx}{\mbox{$\partial_{x}$}}
\ncm{\dxi}{\mbox{$\partial_{x_i}$}}
\ncm{\dxt}{\mbox{$\partial^2_{x}$}}
\ncm{\dxit}{\mbox{$\partial^2_{x_i}$}}
\ncm{\dt}{\mbox{$\partial_{t}$}}
\ncm{\dtt}{\mbox{$\partial_{t}^2$}}
\ncm{\pf}{\mbox{$p_{\phi}$}}
\ncm{\RE}{\mbox{Re}}
\ncm{\IM}{\mbox{Im}}
\ncm{\Tr}{\mbox{tr}}
\ncm{\diag}{\mbox{diag}}
\ncm{\ra}{\rightarrow}
\ncm{\la}{\leftarrow}
\ncm{\dg}{\dagger}
\ncm{\ha}{\hat{a}}
\ncm{\hP}{\hat{P}}
\ncm{\sL}{\sqrt{\Lambda}}
\ncm{\lb}{\overline{\lambda}}
\ncm{\aldot}{\mbox{$\dot{\alpha}$}}
\ncm{\dota}{\mbox{$\dot{a}$}}
\ncm{\dotf}{\mbox{$\dot{\phi}$}}
\ncm{\dfo}{\mbox{$\partial_{\phi_0}$}}
\ncm{\aplt}{ \mbox{}_{\textstyle \sim}^{\textstyle < }     }
\ncm{\apgt}{ \mbox{}_{\textstyle \sim}^{\textstyle > }     }
\ncm{\Oa}{\mbox{$\mbox{O}(a)$}}
\ncm{\Sp}{\hspace{1.0cm}}
\def\be{\begin{equation}}
\def\ee{\end{equation}}
\def\bea{\begin{eqnarray}}
\def\eea{\end{eqnarray}}
\rencm{\thefootnote}{\mbox{\protect{$\fnsymbol{footnote}$}} }
\ncm{\fig}[2]{\begin{figure}[ppp]\caption{\label{#1} #2}\end{figure}}
\ncm{\sect}[1]{\section{#1}\setcounter{equation}{0}}
\ncm{\append}{  \setcounter{section}{1}\setcounter{equation}{0}
          \section*{Appendix}
   \rencm{\theequation}{\Alph{section}.\arabic{equation}}  }
\ncm{\appendA}{  \setcounter{section}{1}\setcounter{equation}{0}
          \section*{Appendix A}
   \rencm{\theequation}{\Alph{section}.\arabic{equation}}  }
\ncm{\appendB}{  \setcounter{section}{2}\setcounter{equation}{0}
          \section*{Appendix B}
   \rencm{\theequation}{\Alph{section}.\arabic{equation}}  }
\ncm{\front}[5]{
\begin{titlepage}
\noindent \Sp\Sp{#1} \hfill {#2}\Sp\Sp\Sp\\
\begin{center}
\vspace{2\baselineskip}
{\Large\bf  #3  } \\
\vspace{3\baselineskip}
\vspace{2\baselineskip}
 #4\\
\vspace{2\baselineskip}
$^1$ HLRZ c/o KFA J\"ulich, P.O. Box 1913, D-5170 J\"ulich, Germany \\
$^2$ Institut f\"ur Theoretische Physik, Universit\"at Bern, \\
Sidlerstrasse 5, CH-3012 Bern, Switzerland

\end{center}
\vfill
{\bf Abstract}\\
 #5
\end{titlepage} }

\begin{document}
\front{May 1992}{HLRZ-92-25/BUTP-92-19}
{Gauge Fixing on the Lattice  \\  without Ambiguity}
{Jeroen C.\ Vink$^1$ and Uwe-Jens Wiese$^2$
 \footnote{on leave from HLRZ c/o KFA J\"ulich, P.O. Box 1913,
 D-5170, Germany, \\ supported by Schweizer Nationalfond} }
{A new gauge fixing condition is discussed, which is (lattice) rotation
invariant, has the `smoothness' properties of the Landau gauge but
can be efficiently computed and is unambiguous for almost all
lattice gauge field configurations.}

\subsection*{Introduction}

In gauge theories gauge fixing is necessary to make the perturbation
expansion well defined. One of the advantages of the lattice regularization of
gauge theories, compared to the continuum formulation, is that gauge
fixing on the lattice is not necessary. This is true if
only gauge invariant operators are considered.
In a growing number of cases, however, it is found advantageous to
also use gauge variant operators which makes it necessary to fix the
gauge even on the lattice.

For instance, it is interesting to study the gauge variant propagators
of quarks and gluons \cite{MaOg87}.
Also when one wants to compare with gauge dependent
results from perturbation theory gauge fixing on the lattice is necessary.
In other applications one leaves
out the strings of gauge parallel transporters
that make operators gauge invariant, like
in point split fermion bilinears or extended glueball operators.
Since this eliminates the fluctuations of the gauge fields contained
in the string, it significantly improves the signal to noise
ratio in a Monte Carlo experiment \cite{weakM}.
To restore gauge invariance one must now
fix to some uniquely defined gauge. Popular choices are the Landau or Coulomb
gauge. An alternative picture justifying this strategy
of leaving out parallel transporters
is that by fixing to the Coulomb gauge, one dresses
the bare charge with its physical gauge boson cloud
instead of squeezing this cloud through a connecting flux string.
For noncompact QED this interpretation of the Coulomb
gauge in terms of a photon cloud can be substantiated \cite{Froh85,PoWi90}.

Let us consider the non-gauge fixed gauge field $U_{\m}(x)\in G$ with gauge
group $G=SU(n)$ or $U(n)$.
Suppose the gauge fixed configuration minimizes a (gauge dependent) function
$R(U)$. An example is the lattice Landau gauge fixing
condition, where
\be
R(U)= \RE \sum_{x,\m}[1- \frac{1}{n} \Tr (U_{\m}(x))].
\label{Lancon}
\ee
The gauge fixing algorithm should provide the transformation $\O$ such that
$R(U^{\O})$ assumes its minimal value, where
\be
U^{\O}_{\m}(x) = \O(x)U_{\m}(x)\O(x+\mh)^{\dg}.
\ee
If the starting configuration is gauge transformed with
$g$, it should find the transformed
\be
\O^g(x)= g_0\O(x)g(x)^{\dg},\;\;
\O^g(x)^{\dg}=g(x)\O^{\dg}(x)g_0^{\dg}, \label{gtrans}
\ee
where the global gauge transformation $g_0$ need not be fixed.
The algorithm should find a unique solution. This means that
for all gauge equivalent initial configurations,
it should find the same gauge fixed configuration
(up to a global gauge transformation $g_0$).
It is clear that this is impossible in principle if the absolute
minimum of $R$ is degenerate. But if this would be the only source
of ambiguity, one would  argue that such an exact degeneracy is
an exceptional situation which will not occur in practice.
However, finding the absolute minimum of
$R$ is extremely difficult and iterative minimalization procedures
usually run into some local minimum. Finding the absolute minimum is a
numerical problem of non-polynomial complexity.

In the continuum limit, the local minima of $R$ correspond to solutions of the
differential Landau condition and the existence of multiple solutions was first
discussed by Gribov \cite{Grib80}.
Also on the lattice we shall call gauge fields for which $R$ has
a (local) minimum Gribov copies. The existence of Gribov copies on the lattice
is demonstrated in \cite{NaPl90,PaPe91}.

In order to remove the Gribov ambiguity it has been proposed to first fix to
a uniquely defined axial gauge \cite{MaOg90}. But even if the
starting gauge field is unique, one can still run into different local minima
by using different gauge fixing algorithms. Moreover the axial gauge breaks
rotation invariance, and some effect of this might be present in the local
minimum that is favored by this particular starting configuration.

Another approach to remove the ambiguity is to include in the gauge fixing
condition the
requirement of averaging over all initial gauge transformations, such that one
averages over a large number of local minima. In practice this procedure is
too expensive. Furthermore it is to be expected
that the set of local minima that will be found
depends on the algorithm used, e.g. it may be different for local
algorithms and for non-local ones like multigrid \cite{HuLa91}.
Finally, averaging over many
local minima re-introduces noise in gauge variant operators
\cite{NaPl90}.

In the following we shall discuss a gauge fixing condition which
can be computed efficiently and unambiguously for all but
a set of exceptional configurations.
It shares desirabel properties, like smoothness
of the gauge fixed configuration, with the Landau-like gauges.
Of course,
it should be noted that we are not solving the Gribov problem
for the usual, local, Landau and Coulomb gauge. We avoid the problem
by constructing a modified gauge condition which is no longer local.
Because of this non-locality
it is not clear if our gauge fixing condition is well suited for
perturbative calculations. If not, this will complicate
a comparison of gauge dependent
results from perturbation theory with nonperturbative lattice
calculations.
Unlike axial gauges our gauge fixing
does not violate global symmetries like e.g. rotation invariance. Therefore
it is well suited for the construction of improved
interpolating fields for hadrons. To project on a definite symmetry sector of
the physical Hilbert space it is essential that the hadron operators transform
appropriately under the various global symmetries. It should be noted that
the transfer matrix formalism requires the operators to be well localized in
euclidean time. This suggests to fix to Coulomb- (rather than Landau-) like
gauges when one has this application in mind.

\subsection*{Laplacian gauge fixing}

For a given gauge field configuration, we want to find a gauge
transformation $\O\in G$
which transforms as in (\ref{gtrans})
and captures the physical `smoothness' of the gauge configurations.
Relaxing the property that $\O\in G$, the above requirements
are reminiscent of finding  covariant
renormalization group transformations or multigrid blocking kernels and
this  may suggest to consider eigenfunctions of the gauge covariant
laplacian $\D(U)$ as in \cite{HSV90,BrRe91,Kalk91}.
The covariant laplacian (up to a minus sign) is defined by
\be
 \D(U)(x,y)^{ab} := \sum_{\m} [    2\d(x-y)\d^{ab} -
  U_{\m}(x)^{ab}\d(x+\mh-y) -
              U_{\m}(y)^{\dg ab}\d(y+\mh-x) ]   \label{deflap}
\ee
and it is a positive (semi) definite operator.
We consider a finite lattice with
lattice spacing $a = 1$ and with periodic boundary conditions.
The eigenfunctions $f^s$ of $\D(U)$ are defined by
\be
 \D(U)(x,y)^{ab}f(y)_b^s = \l^s f(x)_a^s,
\ee
where the $\l^s \ge 0$ are the eigenvalues.
We shall often suppress a summation over repeated indices.
For a non-degenerate spectrum, these eigenfunctions are unique
up to a global phase,
\be
f(x)_a^s\ra \exp(i\h^s) f(x)_a^s.  \label{phasef}
\ee

Under gauge transformations $U\ra U^g$
the eigenfunctions transform as $f(x)_a^s\ra g(x)^{ab}f(x)_b^s$.
This implies that the $GL(n)$ matrix field
\be
M(x)^{ab} := f(x)^b_a    \label{defM}
\ee
has the same transformation behavior as gauge transformations
$\O(x)^{\dg ab}$, cf.(\ref{gtrans}).
The index $b$ in (\ref{defM}) labels a set of $n$ eigenfunctions.
The transformation  behavior of $M$ does not depend on which
$n$ eigenfunctions we choose, but in order to select the smooth
modes in the gauge field, we shall choose the eigenfunctions with
the lowest $n$ eigenvalues. This is in the spirit of refs.
\cite{HSV90,BrRe91,Kalk91} and will be discussed further below.

In general the matrices $M(x)^{ab}$ are not yet in the gauge group $G$.
However, almost all $M$ can be uniquely decomposed as
$M=\O^{\dg} R$ with $\O \in G$ and transforming as in (\ref{gtrans}).
To see this for the case of $G=SU(n)$, we first write $M=WP$ with
$W$ a unitary matrix and $P$ a positive hermitian matrix.
This decomposition is unique if
\be
 P = (M^{\dg}M)^{1/2} =
 S^{\dg}\diag(\m_1^{1/2},...,\m_n^{1/2}) S  \label{defP}
\ee
is invertible, i.e. if $M$ has a non-zero determinant.
The unitary matrix $S$ in (\ref{defP}) diagonalizes $M^{\dg}M$ and
the $\m_a$ are the eigenvalues.
Then $W=MP^{-1}$ and $\O^{\dg}=W\exp(-i\a)$ with
$\exp i\a = \diag(\det(W)^{1/n},...,\det(W)^{1/n})$.
If the gauge group is $U(n)$ the determinant need not be factored out.

More explicitly we define the gauge transformation $\O$ leading to our
Laplacian gauge by
\be
 \O_{\D}(x)^{\dg ab}= f(x)_a^c (P(x)^{-1})^{cb}e^{-i\a(x)}. \label{defO}
\ee
It is clear from the construction that this $\O_{\D}$ indeed transforms
as in (\ref{gtrans}), whereas $P$ and $\a$ are invariant.

The global $U(1)$ transformations of the eigenfunctions (\ref{phasef})
induce global abelian transformations of $\O_{\D}$. This follows
from the observation that $M^{ab}\ra M^{ab}\exp(i\h^b)$ implies
that $S^{ab}\ra \exp(-i\h^a) S^{ab}\exp(i\h^b)$ and consequently
$P^{ab}\ra \exp(-i\h^a) P^{ab}\exp(i\h^b)$ as well. From (\ref{defO})
it then follows that
\be
 \O_{\D}(x)^{ab}\ra \exp(-i\h^a - \overline{\h})\O_{\D}(x)^{ab}, \label{transO}
\ee
where $\overline{\h}=\sum_a \h^a/n$.

In general there is no global $SU(n)$ invariance, only if the lowest
$n$ eigenvalues are degenerate one can show along similar lines as
above, that a global $SU(n)$ invariance is present.

\subsection*{Singularities and ambiguities}

The construction of $\O_{\D}$ outlined above is unambigous for almost
all gauge field configurations: For a given gauge field $U$ the
spectrum of $\D(U)$ is invariant under gauge transformations $g$
and if the spectrum is non-degenerate,
the eigenvalues provide an unambiguous labeling of the
eigenfunctions from which the $\O_{\D}$'s are constructed. In particular,
the $n$ eigenfunctions with the lowest eigenvalues can then
be computed  unambiguously (up to a global phase). The subsequent
decomposition  works if $\det(M(x))\not=0$ for all $x$.

It is clear form this resum\'e that there are two sources of
ambiguity in our construction. If the $(n+1)$-th eigenvalue
is degenerate with the $n$-th (labeling them in ascending order),
the prescription to take the `lowest $n$' is not
well-defined. This means that $\O_{\D}(U)$ becomes ambiguous at those
points in configuration space, where level $n$ and $n+1$ cross.
The level crossing condition fixes one degree of freedom. Hence
the singular manifold is of co-dimension one.
The second source of ambiguity is at points in configuration
space where one or more of the $\det(M(U))$ become zero.
The condition that the complex
determinant vanishes at some point $x$ fixes two degrees of freedom and hence
the singular manifold is of co-dimension two.
In addition, for $SU(n)$ the
determinant has to be factored out. This requires to take the
$n$-th root of it,
which is well-defined only if the determinant is not negative real. This
excludes a singular manifold of co-dimension one.
In the continuum limit it is expected that
these singular sub-manifolds become
Gribov horizons giving configuration space the required non-trivial
topology to incorporate flux sectors and $\q$ vacua \cite{VanB91}.

This discussion shows that
the probability to find singular configurations
in  Monte Carlo simulations on a finite lattice is zero and our
gauge fixing condition is unambiguous for all practical purposes.

\subsection*{Relation to the Landau gauge}

It is an important property of the Landau gauge that the gauge
fixed configuration is smooth in the sense that the link field
$U_{\m}(x)$ has  similar fluctuations as the gauge invariant
plaquette field. Of course there will be localized fluctuations as well as
obstructions due to nontrivial topology (instantons), but on the average $U$
is close to one if the average plaquette is close to one.
By constructing gauge transformations
from the eigenfunctions of the covariant laplacian and
selecting the modes with smallest eigenvalues, one expects that
the gauge fixed configurations are again smooth in the above sense.

To illustrate this we shall first consider pure gauge
configurations, $U_{\m}(x)=g(x)g(x+\mh)^{\dg}$. Here the Landau
gauge fixing condition has the solution $U^{\O}_{\m}(x)=1$.
The eigenfunctions of the laplacian are
gauge transforms of plane waves,
\bea
   f(x)_a^s & = & g(x)^{ab}\exp(ikx), \nonumber \\
   \l^s & = & \sum_{\m}(2\mbox{sin}(k_{\m}/2))^2,
\eea
where the lattice volume is $V=N^4$ and the momenta are
$k_{\m}=2\p n_{\m}/N$, $n_{\m}=0,...,N-1$. The label $s$ is such that
$s = 1,...,n$ corresponds to the $n$ zero momentum modes.
For the pure gauge case the construction
of $\O_{\D}$ is simple and one finds
\be
\O_{\D}(x) = \O_0g(x)^{\dg},
\ee
where $\O_0$ is a global $SU(n)$ transformation which is
not specified by our construction. Note that these are precisely
the gauge transformations that fix $U$ to the Landau gauge.

Since the space dependence of the eigenfunctions for this simple
case comes in as a $U(1)$ factor which drops out in the reduction
to $SU(n)$ we cannot illustrate the importance of taking the
low eigenvalue modes here. With $g\in U(n)$, however, our construction
gives $\O_{\D}(x) = \O_0g(x)^{\dg}\exp(ikx)$, if eigenfunctions
with momentum label $k$ are used. Only for $k=0$ this corresponds to the
Landau gauge, for $|k|>0$ the gauge fixed configuration is less
smooth.

To elucidate the generic case, we consider the function
\be
  E(f) :=  \sum_{x,y} f(x)_a^{\dg}\D(U)(x,y)^{ab}f(y)_b
          \label{defE}
\ee
Minimalizing $E$ under the constraint that
$\sum_x f(x)_a^{s\dg}f(x)_a^t=V\d^{st}$ provides
the eigenfunctions of $\D(U)$ as solutions (normalized to the volume $V$).
After substituting the decomposition (\ref{defO}) in (\ref{defE}) one finds
\be
  \sum_a \l^a = 2n\RE \sum_{x,\m}[1 -
 \frac{1}{n}\Tr (e^{i\a(x)}P(x)\O_{\D}(x)U_{\m}(x)
 \O_{\D}(x+\mh)^{\dg}P(x+\mh)e^{-i\a(x+\mh)})   ]  \label{lamsum}
\ee
This leads one to consider a modified Landau function $R_{\D}$,
\be
R_{\D}(U^{\O}) = \RE \sum_{x,\m}[1 -
\frac{1}{n}\Tr (W_{\m}(x)\O(x)U_{\m}(x)
 \O(x+\mh)^{\dg})].
\label{Lapcon}
\ee
The weight function $W_{\m}(x)$ is gauge invariant and follows from the
eigenfunction construction,
\be
W_{\m}(x) = P(x+\mh)P(x)e^{i(\a(x)-\a(x+\mh))}.
\ee
The minimum of $R_{\D}$ over gauge orbits $\O$ is assumed for $\O = \O_{\D}$,
coming from the eigenfunction construction: $R^{min}_{\D} =
\sum_{a=1}^n \l_a/2n$.
The weight function $W_{\m}(x)$ is a function of all gauge links
$U_{\m}(x)$ and in this sense our gauge condition is non-local.
Clearly $R_{\D}$ reduces to the (local) Landau function
(\ref{Lancon}) if $W_{\m}(x)_{ab} \ra \d_{ab}$.

This shows that our laplacian gauge condition corresponds to finding the
absolute minimum of a modified Landau function (\ref{Lapcon}). Trying to find
the absolute minimum of $R_{\D}$ for a fixed $W_{\m}(x)$ is practically
impossible, but by using the $W_{\m}(x)$ which follows from the eigenfunction
construction, the problem is easily solved.

\subsection*{Discussion}

We have shown that a smooth gauge, similar to the Landau gauge,
can be defined unambiguously on the lattice, except for configurations
which have measure zero in configuration space. In order to find
it for gauge group $SU(n)$ or $U(n)$, one needs to compute the
eigenfunctions corresponding to the $n$ smallest eigenvalues
of the covariant lattice laplacian and make the reduction
(\ref{defO}) to $SU(n)$ or $U(n)$ matrices.
The time consuming part is the computation of the $n$
lowest eigenfunctions.
They  can be computed with the Lanczos algorithm,
or they can be found by inverse iteration (see e.g \cite{Kalk91}),
after the $n$ eigenvalues have been computed with the  Lanczos algorithm.
Since the laplacian is hermitian, this should be feasible even on
large lattices.

The main virtue of our gauge fixing procedure is
the possibility to uniquely specify a smooth Landau-like gauge
in non-perturbative lattice computations without breaking global symmetries
like rotation invariance. At present it is not clear (to us) if laplacian
gauge fixing can also be used in perturbation theory. Even if this is not
possible, laplacian gauge fixed fields may be useful in the construction of
improved hadron or glueball operators.
For this purpose the Wuppertal group \cite{Gues90}
has used the inverse covariant laplacian before,
which is reminiscent of our approach.
A construction based on laplacian gauge fixing,
however,
offers the possibility to interpret results physically
in terms of gauge fixed quark and gluon fields.

\subsection*{Acknowledgement}

We have enjoyed stimulating discussions with Morten Laursen
and Pierre van Baal.

\end{document}